\documentclass[prl, twocolumn]{revtex4-1}
\usepackage{graphicx,epstopdf,mathrsfs,amsmath}
\begin{document}
\author{Soumya Jyoti Banerjee}
\affiliation{Bose Institute, 93/1 Acharya PC Roy Road,Kolkata 700009 India}
\author{Soumen Roy}
\affiliation{Bose Institute, 93/1 Acharya PC Roy Road,Kolkata 700009 India}
%\author{Arising from Y-Y Liu, J-J Slotine \& A-L Barabasi Nature {\bf 473}, 167 - 173 (2011)}

\title{Key to Network Controllability\\~\\
{\small {\sf Arising from YY Liu,  JJ Slotine \& AL Barab\'asi, Nature {\bf 473}, 167 - 173 (2011)}}}
\maketitle

%\begin{abstract}
Liu {\it et. al.} recently proposed a minimum number of {\em driver nodes}, $N_D$, needed to obtain full structural controllability over a directed network~\cite{Liu}. Driver nodes are unmatched nodes, from which there are directed paths to all matched nodes. Their most important assertion is that a system's controllability is to a great extent encoded by the underlying network's degree distribution, $P(k_{in}, k_{out})$.  Is the controllability of a network decided almost completely by the immediate neighbourhood of a node, while, even slightly distant nodes play no role at all? Motivated by the above question, in this communication, we argue that an effective understanding of controllability in directed networks can be reached using distance based measures of closeness centrality $(CC)$ and betweenness centrality $(BC)$ and may not require the knowledge of  local connectivity measures like in-degree and out-degree. 
%\end{abstract}
%\maketitle

%%%%%%%%%%%%%%%%%%%%%%%%%%%%%%%%%%%%%%%%%%%%%%%%%%%%%%%
\begin{figure}[t]
\vspace{-0.26in}%
\hspace{-0.35in}\includegraphics[width=1.1\columnwidth, height = 90mm]{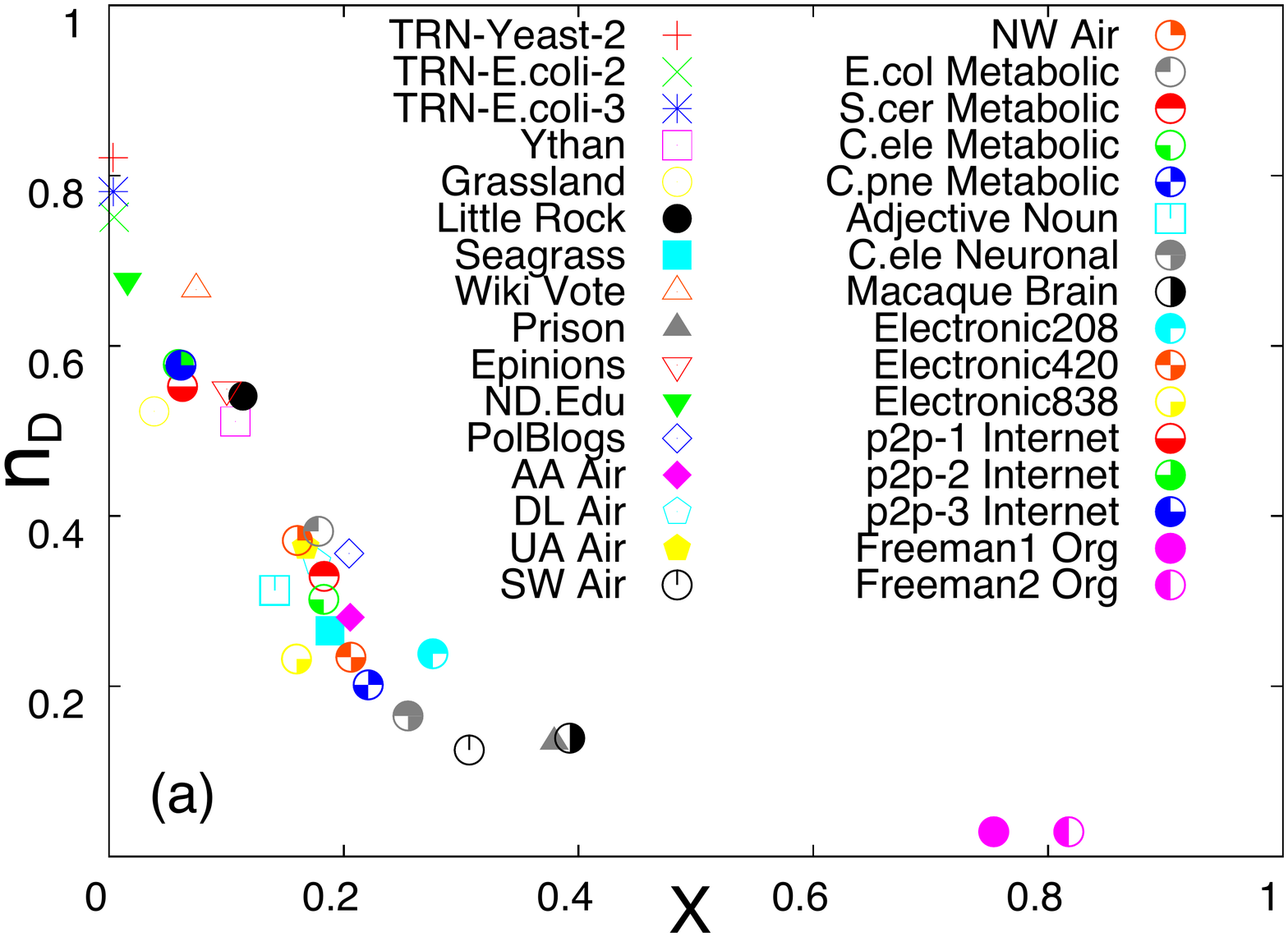}\\%
\vspace{-0.46in}%
\hspace{-0.42in}\includegraphics[width=1.12\columnwidth, height = 65mm]{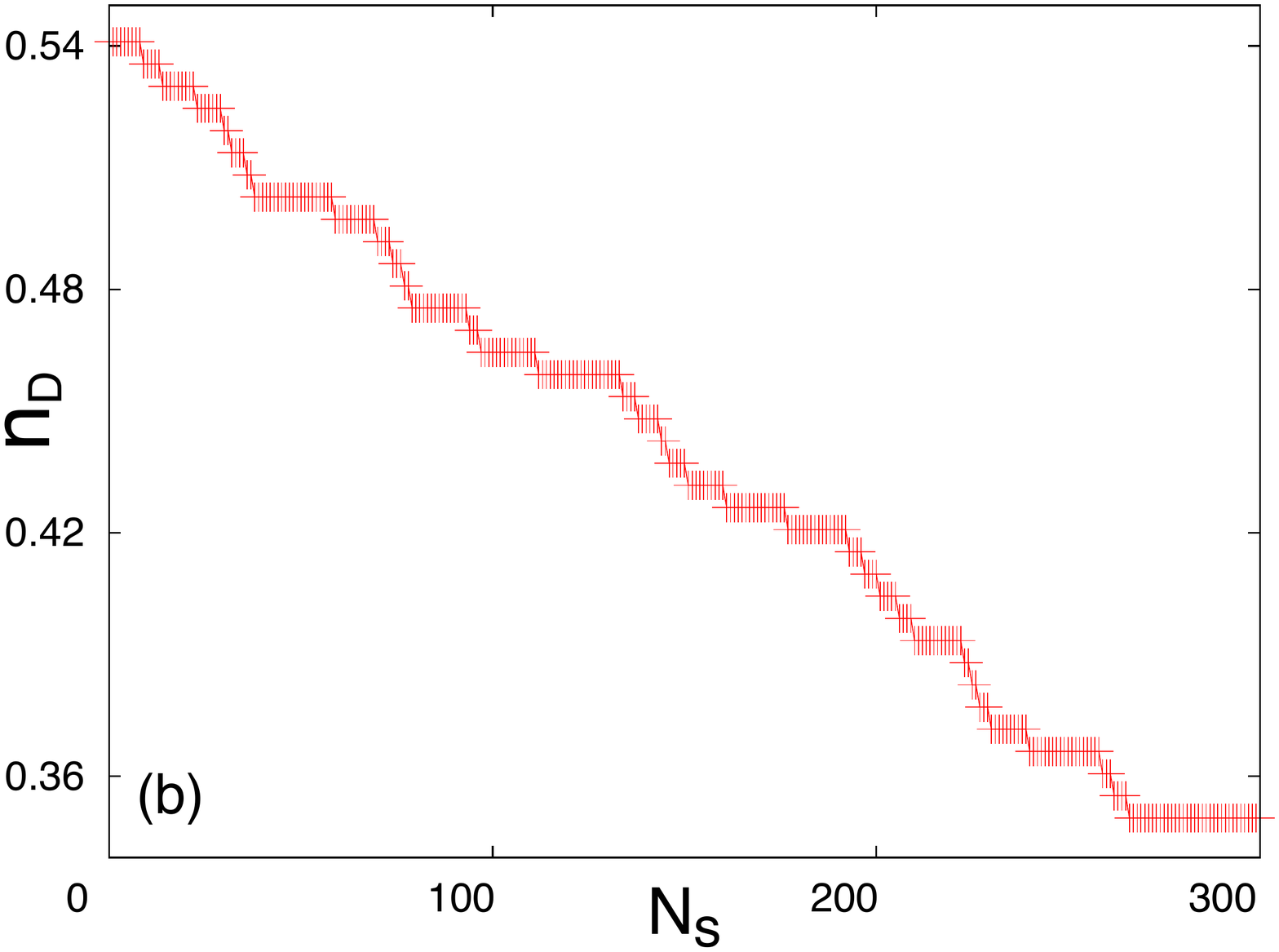}%
\vspace{-0.22in}%
\caption{(a) Change in controllability, $n_D$, versus $X = <{\mathscr C}> +<{\mathscr B}>/<{\mathscr C}>$ %$X({\mathscr {C, B}}) = <{\mathscr C}> +<{\mathscr B}>/<{\mathscr C}>$
 for $32$ different real world networks with different degree distributions from ~\cite{Liu,airline,dir-url}. (b) Change in $n_D$ for Little Rock food web (which showed maximum deviation between actual  and expected values~\cite{Liu}) versus $N_S$ (number of swaps). Swapping procedure is detailed in {\em Methods}. Only $12\%$ of edge swaps lead to decrease of $n_D$ by $35\%$ whereas Ref.~\cite{Liu} predicts $n_D$ to remain almost constant for such swaps.} 
\label{fig1}
\end{figure}
%%%%%%%%%%%%%%%%%%%%%%%%%%%%%%%%%%%%%%%%%%%%%%%%%%%%%%%

Consider, for example, the case of two $N=4$ node networks: a chain graph, ${\mathscr {G}}_1$, and a rather densely connected   graph, ${\mathscr {G}}_2$, constituted by the set of edges   ${\mathscr {E}}_1= \{(1,2), (2,3), (3,4)\}$  and  ${\mathscr {E}}_2 = \{(1,2), (1,3), (1,4), (2,3), (2,4), (3,4)\}$ respectively. It is apparent that graphs {\em like}  ${\mathscr {G}}_1$and  ${\mathscr {G}}_2$ would have very different $P(k_{in}, k_{out})$ and degree correlations. Therefore,  ${\mathscr {G}}_1$ and  ${\mathscr {G}}_2$ are expected to have different controllability, $(n_D = N_D/N)$, values following Ref.~\cite{Liu}. However,  $n_D = 1/4$, for both ${\mathscr {G}}_1$ and  ${\mathscr {G}}_2$. It is also notable that there exists some difference between the actual and expected values of $n_D$ for a number of degree preserved random graphs. Indeed, for food webs, metabolic and neuronal networks, this difference is significant~\cite{Liu, correlations}. 

%%%%%%%%%%%%%%%%%%%%%%%%%%%%%%%%%%%%%%%%%%%%%%%%%%%%%%%
Degree reflects information about the immediate neighbourhood of a node. 
However, $CC$ signifies a node's potential to choose good control paths passing through it.
%However, $CC$  indicates a node's potential to distinguish between paths emerging from it on basis of controllability. 
Again, a node with high $BC$ could be rather distant from the node in question but could connect it to a matched or unmatched node. Thus, we are immediately led to investigate the important role that $CC$ and $BC$ should play in deciding controllability. 
%closeness and betweenness centralities of the nodes of a network.  
We propose that controllability should obey a functional relation, $n_D = F(X({\mathscr {C, B}} ))$, where, ${\mathscr C}=(C_1,C_2,...,C_N)$ and ${\mathscr B}=(B_1,B_2,...,B_N)$, for a network of $N$ nodes. $C_i$ and $B_i$, where $i\in\{1,2,3,...,N\}$; are defined in {\em Methods}. Herein, we examine the dependence of $n_D$ on $X$, where $X$ is a very simple function: 
\begin{equation}
X({\mathscr {C, B}} ) = 
\begin{cases}
<{\mathscr C}>+<{\mathscr B}>/<{\mathscr C}>, & \text{if } <{\mathscr B}> \ne 0 \\
<{\mathscr C}>, & \text{if }<{\mathscr B}> = 0
\end{cases}
\label{function1}
\end{equation}
$<{\mathscr C}> \ne 0$ in a connected network and hence $X({\mathscr {C, B}} )$ is always well-defined. 
 $<{\mathscr C}>$ and $<{\mathscr B}>$ denote the average (over all nodes) of closeness and betweenness centralities respectively of a network.  
We now focus on the limiting cases of best and worst controllability. The former is observed in directed, fully connected graphs and chains and the latter in directed star graphs. However, low $X ({\mathscr {C, B}})$ in star networks lead to high $n_D$ whereas high $X ({\mathscr {C, B}})$ in completely connected  graphs and chains lead to low $n_D$. 

 As apparent from Fig.\ref {fig1}(a), assuming Function~\ref{function1} involving $<{\mathscr C}>$ and $<{\mathscr B}>$ presents a coherent picture of controllability.  $n_D$ decreases, with increase of $X ({\mathscr {C, B}})$. The term $<{\mathscr B}>/<{\mathscr C}>$ reflects the interaction among betweenness and closeness in deciding controllability. {\em The value of} $X ({\mathscr {C, B}})$ {\em for a given network furnishes an estimate of it's controllability, without resorting to maximum matching which is computationally expensive}. Also, $X ({\mathscr {C, B}})$ seems to act as a good index for distinguishing various kinds of biological networks: very low ($ \approx 0.003$), low ($\approx0.2$) and intermediate ($\approx0.33$) values of $X ({\mathscr {C, B}})$ correspond to transcriptional, metabolic and neuronal networks respectively. $X ({\mathscr {C, B}})$ also illustrates why dense graphs are easier to control than sparser ones.  Fig.~\ref {fig1}(b) shows that controllability can be changed even within a given network by using $X ({\mathscr {C, B}})$ and without taking recourse to degree correlations~\cite{correlations}. 

We wish to emphasize that (i) we have considered only the first standardised moment of ${\mathscr {C}} $ and ${\mathscr {B}}$ for simplicity, and, (ii) the most efficient strategy to destroy network controllability should naturally exploit $CC$ and $BC$.  We are addressing (Banerjee and Roy, in preparation), the former by incorporating higher standardised moments of metrics which are known to convey a clearer picture in networks~\cite{ssb,pre091,epl091}, and, the latter by using $CC$ and $BC$ rather than control centrality~\cite{control-centrality}.  In summary, we present an understanding of controllability in directed networks using distance based measures of  closeness and betweenness centralities, which can be achieved without using local connectivity measures like in-degree and out-degree. 

%%%%%%%%%%%%%%%%%%%%%%%%%%%%%%%%%%%%%%%%%%%%%%%%%%%%%%
\section{Methods summary}
\label{sec:methods} 

{\em Closeness centrality} of a node, ${i}$, is the inverse of the sum of its distance $d(i,j)$, to all other nodes, ${j}$. Mathematically, it is defined as:
\begin{equation}
C_{i} = \frac{N-1}{\displaystyle \sum  _{j=1}^{N} d({i}, {j}) }
\end{equation}
{\em Betweenness centrality} of a node, $i$, refers to the ratio of the number of shortest paths, $\sigma_{st}(i)$, that pass through $i$ to the total number of shortest paths, $\sigma_{st}$, existing from node, $s$, to node, $t$, in the network. Mathematically, it is defined as:
\begin{equation}
B_i = \displaystyle\sum_{s \not= i \not= t} \frac{\sigma_{st}(i)}{\sigma_{st}}
\end{equation}
{\em Edge swapping procedure:} a pair of edges is swapped only if it increases $X({\mathscr {C, B}} )$ while preserving in and out degree, (and hence total degree), of every node involved. Thus $P(k_{in},k_{out})$ for a network remains unchanged. 
%%%%%%%%%%%%%%%%%%%%%%%%%%%%%%%%%%%%%%%%
\section{Acknowledgement}
SJB thanks CSIR for financial support. 
%%%%%%%%%%%%%%%%%%%%%%%%%%%%%%%%%%%%%%%%
 
\end{document}